\documentclass[11pt]{article} 

\usepackage{pgf}
\usepackage{graphicx}
\usepackage{url,booktabs,multirow}
\usepackage{hhline}
\usepackage{amsmath,amsfonts,amssymb,amsbsy,amsthm,bm, bbm, mathrsfs}
\usepackage{dsfont} 
\usepackage{float}
\usepackage{algorithm} 
\usepackage{algpseudocodex} 
\usepackage{nicefrac}
\usepackage{subfigure}
\usepackage{color}
\usepackage{times}
\usepackage{multicol}
\usepackage{setspace}
\usepackage{parskip}
\usepackage[utf8]{inputenc}
\usepackage{longtable}
\usepackage[labelfont=bf,
width=\textwidth, justification=justified]{caption} 
\usepackage{float}
\usepackage[section]{placeins}
\usepackage{mathtools}
\usepackage{tabularx}
\usepackage{layouts} 
\usepackage{csquotes}
\usepackage{easyReview}
\usepackage[colorlinks]{hyperref}
\usepackage{cleveref}
\usepackage{orcidlink}
\usepackage{authblk}
\usepackage[english]{babel}

\usepackage[nolist]{acronym}
\numberwithin{equation}{section}

\definecolor{bleudefrance}{rgb}{0.19, 0.55, 0.91}{\tiny}

\usepackage[
backend=biber,
style=apa,
sortlocale=de_DE,
natbib=true,
url=true, 
doi=true,
eprint=false,
uniquelist = false, %when more than one et al is citation from same author happens to be in the same year..
]{biblatex}

\begin{document}
\begin{acronym}
	\acro{AIC}{Akaike information criterion}
	\acro{BLUE}{best linear unbiased estimator}
	\acro{KDE}{kernel density estimation}
	\acro{GRSST}{Groß–Rendtel-Schmid–Schmon–Tzavidis}
	\acro{AGRSST}{augmented GRSST}
	\acro{SEM}{stochastic expectation–maximization}
	\acro{SAE}{small area estimation}
	\acro{PDF}{probability density function}
	\acro{PMF}{probability mass function}
	\acro{RMISE}{root mean integrated squared error}
	\acro{MIAE}{mean integrated absolute error}
	\acro{EM}{expectation-maximization}
	\acro{MCEM}{Monte-Carlo EM}
	\acro{NUTS}{Nomenclature of units for territorial statistics}
	\acro{SEM}{stochastic EM}
	\acro{Destatis}{Federal Statistical Office of Germany}
	\acro{GAMM}{generalized additive mixed model}
	\acro{NDVI}{normalized difference vegetation index}
	\acro{NIR}{near-infrared radiation}
	\acro{MODIS}{moderate resolution imaging spectroradiometer}
\end{acronym}

\title{Density Estimation from Aggregated Data with Integrated Auxiliary Information: Estimating Population Densities with Geospatial Data}

\author[1]{Michael Mühlbauer  \orcidlink{0009-0004-1192-3238}}

\author[1]{Timo Schmid  \orcidlink{0000-0002-7217-2501} \thanks{timo.schmid@uni-bamberg.de}}

\affil[1]{Institute of Statistics, University of Bamberg, Bamberg, Germany}

\date{\small Version: Preprint 1.0 (05.08.2025)}

\maketitle
\begin{abstract}
	\noindent 
	Density estimation for geospatial data ideally relies on precise geocoordinates, typically defined by longitude and latitude. However, such detailed information is often unavailable due to confidentiality constraints. As a result, analysts frequently work with spatially aggregated data, commonly visualized through choropleth maps. Approaches that reverse the aggregation process using measurement error models in the context of kernel density estimation have been proposed in the literature. From a methodological perspective, we extend this line of work by incorporating auxiliary information to improve the precision of density estimates derived from aggregated data. Our approach employs a correlation-based weighting scheme to combine the auxiliary density with the estimate obtained from aggregated data. We evaluate the method through a series of model-based simulation scenarios reflecting varying conditions of auxiliary data quality. From an applied perspective, we demonstrate the utility of our method in two real-world case studies: (1) estimating population densities from the 2022 German Census in Bavaria, using satellite imagery of nighttime light emissions as auxiliary data; and (2) analyzing brown hare hunting bag data in the German state of Lower Saxony. Overall, our results show that integrating auxiliary information into the estimation process leads to more precise density estimates.
	\medskip
	
\end{abstract}

\section{Introduction}
\setcounter{page}{1}
\label{intro}

Understanding the spatial distribution of phenomena is relevant across various fields such as public health (see e.g., \citet{yang2022}, who explore clusters of gastrointestinal tumors), urban planning (see e.g., \citet{brandao2018} who estimate the distribution of building renovations in Lisbon) and environmental monitoring, where \citet{lin2010} provides an example of identifying soil pollution hotspots in the Changhua county of central Taiwan. A widely applied non-parametric method for estimating spatial distributions is \ac{KDE} (see \citet{silverman2018} for a general introduction and \citet{rosenblatt1956} and \citet{parzen1962} for the foundational work on the method), which traditionally requires precise data points, such as exact 2D geocoordinates comprised of longitude and latitude values for bivariate spatial analysis. Such precise data is not always available to analysts due to reasons such as the data collection method or post hoc spatial aggregation implemented to address confidentiality concerns. For example, vaccination data for common diseases such as Tick-borne encephalitis (TBE) and influenza are only published at the level of the German \ac{NUTS}-3 \citep{rieck2020}. Similarly, the European brown hare (colloquially also referred to as "jackrabbits") hunting bag data used in this paper is also only recorded at the \ac{NUTS}-3 district level. The \ac{GRSST} estimator \citep{gross2016b}, which is detailed in the next section, is an approach designed for data scenarios characterized by these limitations. The \ac{GRSST} estimator builds upon the core idea of iteratively enriching a flat pilot \ac{KDE} estimate with aggregated information, described in \cite{gross2016b} and extends it to the bivariate context of spatial data. Subsequent literature includes \citep{gross2020}, which demonstrates how the \ac{GRSST} can be used to transform area aggregates between non-hierarchical area systems. In \citet{rendtel2018}, the algorithm is applied in the context of open data to improve the estimation of local childcare needs in Berlin. Later, \citet{rendtel2021} utilize the concept to enhance the visualization of COVID-19 incidence clusters.

An inherent characteristic of the \ac{GRSST} estimator, already identified in the original work by \citet{gross2016b}, is its dependence on the geographical size of the input data areas. Large areas tend to produce largely homogeneous density plateaus on the insides away from their borders, as there is no input information to support a more accurate estimate within those areas, apart from the aggregated information. This implies that, ceteris paribus, the quality of the estimate decreases with fewer and larger input areas. Depending on the context and the underlying distribution, the coarseness of the input areas might be negligible, but when looking e.g. at the \ac{NUTS}-3 district level and the distribution of the human population, accepting a lack of information on the distribution inside of the districts intuitively comes with density estimates of decreased quality. Motivated by the spirit of \ac{SAE} (see \citet{rao2015} for a comprehensive overview), we try to address this shortcoming through the use of auxiliary data. Because oftentimes, analysts may find themselves with auxiliary data that carries some informational value regarding the actual density of interest. For instance, population density tends to correlate with satellite imagery of nighttime light emissions \citep{sutton2001}. Similarly, in the context of ecological data where the distribution of a certain animal species is only recorded per large habitat area, detailed vegetation maps could provide valuable auxiliary information about potentially preferred microhabitats within those broader zones. 

At this point the natural question arises on how to exploit such auxiliary information in order to improve \ac{GRSST} estimate. To achieve this, we propose an additional step to the \ac{GRSST} algorithm: forming a convex combination of an auxiliary density, derived from auxiliary data, and the standard \ac{GRSST} density estimate. The weighting for this combination is based on the correlation calculated between the auxiliary density, aggregated to the spatial level of the GRSST input data, and the GRSST input data itself. We refer to this augmented approach as \ac{AGRSST}. A special case of the \ac{AGRSST} arises when the auxiliary density is weighted with 1; we shall call this special case \ac{AGRSST}1. This version essentially relies fully on the auxiliary density and benchmarks it against the GRSST input data. In this paper, we first recall the \ac{GRSST} algorithm and then introduce our extension, \ac{AGRSST}. Subsequently, we investigate the performance of the \ac{GRSST}, \ac{AGRSST}, and \ac{AGRSST}1 in a model-based simulation study and later in an evaluation study based on real-world German population data from the 2022 Census. We also apply the \ac{AGRSST} to brown hare hunting bag data in the German state of Lower Saxony, using remote sensing \ac{NDVI} data to construct an auxiliary density. Our results indicate that the \ac{AGRSST} can be a valuable addition to the exclusive use of the \ac{GRSST}; however, its strong reliance, especially in the \ac{AGRSST}1 case, necessitates a sound theoretical justification for the auxiliary density. The structure of the paper is as follows: In \Cref{sec:methodology}, we revisit the \ac{GRSST} and propose the additional \ac{AGRSST} step. In \Cref{sec:simulation}, we evaluate our method in a simulation study. \Cref{sec:evaluation} presents an evaluation study that applies the \ac{AGRSST} in the context of population density estimation. In \Cref{sec:application}, we apply our method to a real-world scenario, estimating the density of the brown hare hunting bag. Finally, in \Cref{sec:conclusion}, we conclude our paper with a discussion and final remarks, pointing towards possible future research directions.

\section{Methodology}
\label{sec:methodology}

\subsection{The Groß–Rendtel–Schmid–Schmon–Tzavidis Estimator}
\label{GRSST}

The \ac{GRSST} estimator was first introduced in \citet{gross2016b} to estimate bivariate densities based on rounded geocoordinate data of ethnic minorities and aged people in Berlin. The estimator is based on a measurement error model which allows the derivation of an iterative procedure reminiscent of the \ac{SEM} algorithm \citep{celeux1996} combined with \ac{KDE} in each iteration. In their original work, \citet{gross2016b} propose the \ac{GRSST} estimator to address the case of rounded geocoordinates, where precise geocoordinates are rounded to the center points of the rectangles in which they fall. The following formulation of the \ac{GRSST} estimator is slightly more general, describing the rounding process as the association of a geocoordinate with an area. It is important to note that this formulation is not novel but rather a slight variation of the framework already laid out in \citet{gross2020}. The \ac{GRSST} algorithm is implemented in the \texttt{R} package \texttt{Kernelheaping} (\citet{gross2022}, see \citet{gril2025} for a detailed tutorial).

We assume a geographical map as the sample space $\Omega = \left \{(\ell^{o}, \ell^{a}) \in \mathbb{R}^2 \right \}$ for the random variable $Z$ of the latent (unobserved) geocoordinates $z_i = (\ell^{o}_i, \ell^{a}_i)$, where $\ell^{o}_i$ and $\ell^{a}_i$ are geographical longitude and latitude values and the index $i \in \{1, 2, \ldots, n\}$ refers to the geocoordinates that constitute the total sample size $n$. We employ $\boldsymbol{\ell} = (\ell^{o}, \ell^{a})$, to denote a geocoordinate location independently of the underlying random variable. Further, $Z$ is assumed to follow a continuous \ac{PDF}

\begin{equation}
	\label{eq:z}
	Z \overset{iid}{\sim} P_Z(\boldsymbol{\ell}).
\end{equation}

Many applications of the \ac{GRSST} estimator assume that the entire population is sampled, so $n = N$. We also postulate a random variable $X(Z)$, which maps $Z$ to a set of disjoint sets of geocoordinates, so that 
$X = f(Z): \Omega \rightarrow \left\{A_{1}, .., A_{d}, .., A_{D} \right\}$, where $\Omega = \cup_{d = 1}^{D}A_{d}$. Here, $A_{d}$ denotes the set of all possible geocoordinates that fall inside of the $d$-th geographical area, and $x$ represents a particular outcome of the random variable $X$, where this outcome is one of the defined geographical areas $A_{d}$. For example, if $\Omega$ represents the map of Germany and the $D$ geographical areas are the German \ac{NUTS}-1 regions, then, since the \ac{NUTS}-1 level consists of the 16 federal states, $D = 16$. In this case, one specific $A_d$ would be the set containing all geocoordinates within the state of Bavaria. Consequently, the random variable $X$, representing the observed geographical area, follows a multinomial \ac{PMF} with one trial and $D$ categories.

\begin{equation}
	\label{eq:x}
	X \overset{iid}\sim P_{X}(\boldsymbol{\ell}) = \mathcal{M}_{D, 1}(\boldsymbol{\ell}),
\end{equation}

which can be evaluated at the geocoordinate level using the map $f(Z)$. The conditional probability distribution of the area $X$, given a specific latent geocoordinate $Z$ located at $\boldsymbol{\ell}$ and denoted as $P_{X|Z}(\boldsymbol{\ell})$, follows a Dirac distribution:

\begin{equation}
	\label{eq:dirac}
	P_{X|Z}\left( \boldsymbol{\ell} \right) = 
	\begin{cases}
		1 & \text{if } \boldsymbol{\ell} \in A_{d}, \text{ with } x = A_{d} \\
		0 & \text{otherwise.} \\
	\end{cases}
\end{equation}
We can combine these three expressions in Bayes' theorem and obtain

\begin{equation}
	\label{eq:bayes}
	P_{Z|X}(\boldsymbol{\ell}) = \frac{P_{X|Z}\left( \boldsymbol{\ell} \right) P_Z(\boldsymbol{\ell})}{P_{X}(\boldsymbol{\ell})}.
\end{equation}

This is a measurement error model for the density of the exact, unknown geocoordinate $z_{i}$, given the observed area $x_{i}$ in which it is located. We draw attention to the fact that $P_{X|Z}\left( \boldsymbol{\ell} \right) = \mathbbm{1}_{\boldsymbol{\ell} \in A_d}(\boldsymbol{\ell})$, which allows us to rewrite \cref{eq:bayes} as

\begin{equation}
	\label{eq:propto}
	P_{Z|X}(\boldsymbol{\ell}) \propto P_{X|Z}\left( \boldsymbol{\ell} \right) P_Z(\boldsymbol{\ell}) = \mathbbm{1}_{\boldsymbol{\ell} \in A_d}(\boldsymbol{\ell}) P_Z(\boldsymbol{\ell}).
\end{equation}

Here, we can see that the introduction of $x_{i}$ via Bayes' theorem essentially limits the possible draws of a geocoordinate $z_{i}$ to the area $A_{i}$ in which it is known to be located. 

The core idea of the \ac{GRSST} estimator (formally outlined in \Cref{alg:GRSST}) is to begin with a pilot estimate $P_{Z}^{(0)}(\boldsymbol{\ell})$, incorporate the information contained in $x_{i}$ by sampling according to \cref{eq:bayes} for all $x_{i}$, to update to a new $P_{Z}^{(l)}(\boldsymbol{\ell})$ (the "E-step"), and minimize the \ac{RMISE} using \ac{KDE} (the "M-step"). This process is repeated until convergence. Note that the superscript $(l)$ denotes the iteration number, which runs from $0$ to $M = L + B$, where $B$ is the number of burn-in iterations and $L$ the number of additional iterations. This algorithm closely resembles the classic \ac{EM} algorithm \citep{dempster1977}; however, it is important to emphasize several key differences, which we discuss below.

\begin{algorithm}[H]
	\caption{Groß–Rendtel–Schmid–Schmon–Tzavidis Estimator}
	\label{alg:GRSST}
	\begin{algorithmic}[1] % The number 1 means numbered lines
		\State Start with a pilot estimate of $P_{Z}(\boldsymbol{\ell})$, $\hat{P}_Z^{(0)}(\boldsymbol{\ell})$ by estimating a \ac{KDE} with a large bandwidth $\mathbf{H}$.
		\vspace{5pt}  % Adds vertical space between steps
		\State Evaluate $\hat{P}_Z^{(l-1)}(\boldsymbol{\ell})$ from the last iteration on a fine grid $\mathbf{G} \subset \Omega$ of geocoordinates (grid points) denoted $\boldsymbol{g}_j$, with index $j \in \{1, 2, \ldots, J\}$.
		\vspace{5pt}  % Adds vertical space between steps
		\State S-Step: For all $i$ sample from $\mathbf{G}$ using the corresponding value of $P_{X|Z}\left( x_{i} | \boldsymbol{\ell} \right) \hat{P}_Z^{(l-1)}(\boldsymbol{\ell}) + c$ at each geocoordinate as a sampling weight, to obtain samples from $\hat P_{Z|X}^{(l)}\left(\boldsymbol{\ell} \right)$. The realizations of $x_i$ are given by the data. The constant $c$ acts as an additional smoothing parameter and is set to $1 \cdot 10^{-10}$ in the \texttt{Kernelheaping} package.
		\vspace{5pt}  % Adds vertical space between steps
		\State M-Step: Estimate a new bandwidth $\mathbf{H}$ using the samples from step 3, and use the samples and the new $\mathbf{H}$ to compute the new KDE estimate of the current iteration, $\hat{P}_Z^{(l)}(\boldsymbol{\ell})$.
		\vspace{5pt}  % Adds vertical space between steps
		\State Repeat steps 2–4 for $B$ (burn-in iterations) + $L$ (additional iterations) times.
		\vspace{5pt}  % Adds vertical space between steps
		\State Discard the $B$ burn-in density estimates and obtain the final density estimate of $P_Z(\boldsymbol{\ell})$ by averaging the remaining $L$ density estimates $\hat{P}_Z(\boldsymbol{\ell})$ on the evaluation grid $\mathbf{G}$. The final estimate is called $P_{\text{\tiny GRSST}}(\boldsymbol{\ell})$.
	\end{algorithmic}
\end{algorithm}

It is crucial to observe that the full set of geocoordinates drawn in step 3 of \Cref{alg:GRSST} is, depending on the granularity of $\mathbf{G}$, approximately equivalent to draws from $P_Z^{(l)}(\boldsymbol{\ell})$. For each individual draw, $x_i$ is nonstochastic; the scaling by the marginal likelihood $P_{X}(\boldsymbol{\ell})$ in the denominator of \cref{eq:bayes} is accounted for by the relative empirical frequencies of the $x_{i}$ in the sample. This is especially true in the case where $n = N$.

As already touched upon above, we can observe familiar structures from the \ac{EM} algorithm in the \ac{GRSST} algorithm, but there are significant differences, which we will highlight after a brief introduction to the core components of the \ac{EM} family of algorithms. The \ac{EM} algorithm assumes a fully specified parametric model $P_{Z, X}(z,x \mid \boldsymbol{\theta})$, with a latent variable $Z$, an observed variable $X$, and a vector of parameters $\boldsymbol{\theta}$. Given the parameters $\boldsymbol{\theta}^{(l)}$ of the current iteration $l$, the E-Step typically constructs a function

\begin{equation}
	Q\left(\boldsymbol{\theta} \mid \boldsymbol{\theta}^{(l)}\right) = 
	\int_{\Omega_{Z}} \log \left[ P_{X, Z\mid \boldsymbol{\theta}}(x,z)\right] P_{Z \mid X, \boldsymbol{\theta}^{l}}(z\mid x) dz=
	\mathbb{E}_{Z \mid X, \boldsymbol{\theta}^{(l)}} \left[\log \left[ P_{X, Z \mid \boldsymbol{\theta}}(x,z)\right]\right],
\end{equation}

which is a lower bound of the log-likelihood $\log \left[ P_{X \mid \boldsymbol{\theta}}(x)\right]$. The M-Step maximizes $Q(\boldsymbol{\theta} \mid \boldsymbol{\theta}^{(l)})$ with respect to $\boldsymbol{\theta}$ and thereby provides a new estimate $\boldsymbol{\theta}^{(l+1)}$ for the E-step of the next iteration. The resulting sequence $\boldsymbol{\theta}^{(0)} \ldots \boldsymbol{\theta}^{(M)}$ monotonically increases towards local maxima or the global maximum of $\log \left[ P_{X \mid \boldsymbol{\theta}}(x)\right]$. Convergence properties are studied in more detail by \citet{wu1983}. Note that $P_{Z \mid X, \boldsymbol{\theta}^{(l)}}(z \mid x)$ can be interpreted as a weight for each data point, describing the association of each value of the observed value $x$ with each possible value $z$ given the current parameter estimate $\boldsymbol{\theta}^{(l)}$. These weights enable the appropriate consideration of the current estimate of the distribution of the latent variable $z$ in the subsequent M-Step. For some scenarios, the integration over $z$ in the E-step is prohibitively difficult or impossible. The Monte-Carlo \ac{EM} algorithm, introduced by \citet{wei1990}, approximates $Q\left(\boldsymbol{\theta} \mid \boldsymbol{\theta}^{(l)}\right)$ by

\begin{equation}
	\label{eq:q_mcem}
	Q\left(\boldsymbol{\theta} \mid \boldsymbol{\theta}^{(l)}\right) \approx \frac{1}{T} \sum_{t = 1}^{T} \log \left[ P_{X, Z\mid \boldsymbol{\theta}}(x, \tilde z_{t})\right],
\end{equation}

where $z_{t}$ denotes samples drawn from the current estimate of $P_{Z \mid X, \boldsymbol{\theta}^{(l)}}(z \mid x)$. So, instead of weighting with the current value of $P_{Z \mid X, \boldsymbol{\theta}^{(l)}}(z \mid x)$, $T$ many pseudo-samples of $z$, denoted $\tilde z_{t}$, are drawn for each observed $x$, and the average joint log-likelihood is computed. A special case of the \ac{MCEM} algorithm for a finite mixture model arises for $T = 1$ and is called the \ac{SEM} algorithm \citep{celeux1985}. The M-Step is replaced by the stochastic step (S-Step). The randomness introduced in the S-step causes the sequence of parameter estimates $\boldsymbol{\theta}^{(l)}$ to not converge monotonically towards the closest saddle point, plateau, or local maximum. \citet{celeux1996} argue that the sequence $\boldsymbol{\theta}^{(l)}$ is an irreducible, time-homogeneous Markov chain when $P_{Z \mid X, \boldsymbol{\theta}^{(l)}}(z \mid x)$ is positive for almost every $\boldsymbol{\theta}$ and $z$. If this sequence can be shown to be ergodic, it will converge to the unique stationary probability distribution $\phi$ of the Markov chain. A natural candidate for an estimator of $\boldsymbol{\theta}$ is the average of the last $L$ draws from the chain; see e.g., \citet{nielsen2000}.

Coming back to the \ac{GRSST} algorithm, we observe that, disregarding the logarithm in \cref{eq:q_mcem}, step 3 in \Cref{alg:GRSST} is comparable to the S-Step in the \ac{SEM} algorithm. Using the model's state from the previous iteration, $\hat{P}_Z^{(l-1)}(\boldsymbol{\ell})$, the algorithm completes the dataset by drawing samples from $P_{Z|X}^{(l)}(\boldsymbol{\ell}| x_i) \propto P_{X|Z}\left( x_{i} | \boldsymbol{\ell} \right) \hat{P}_Z^{(l-1)}(\boldsymbol{\ell})$. The model defined in \cref{eq:x}--\cref{eq:bayes} does not make explicit parametric assumptions about the distribution of $Z$. Instead of producing a sequence of parameter estimates $\boldsymbol{\theta}^{(l)}$, the algorithm generates a sequence of \ac{KDE} estimates, $\hat{P}_Z^{(l)}(\boldsymbol{\ell})$. Therefore, the draws in the \ac{GRSST}'s S-Step differ from the classic \ac{SEM} because $P_{Z|X}(\boldsymbol{\ell}| x_i)$ is constructed using last iteration's $\hat{P}_Z(\boldsymbol{\ell})$, unlike last iteration's $P_{Z \mid X, \boldsymbol{\theta}}(\boldsymbol{\ell})$ in the classic \ac{SEM}. This general absence of parametric assumptions also affects the \ac{GRSST}'s M-Step. Instead of maximizing the (average, approximate) joint likelihood as in the \ac{EM}, \ac{MCEM}, or \ac{SEM} algorithms, the \ac{GRSST} algorithm minimizes the asymptotic mean integrated squared error of the kernel density estimator. The replacement of likelihood maximization with a surrogate function is a concept already applied in several generalizations of the \ac{EM} algorithm (see e.g., \citet{elashoff2004}). \citet[][ch. 7.7]{mclachlan2007} provide an overview of the MM algorithm, which stands for Majorization-Minimization or Minorization-Maximization algorithm, a family of algorithms more general than the \ac{EM}. The MM algorithm also allows for minimization in its second step (M-Step), and depending on the underlying model, a host of different surrogate functions can be constructed using relationships such as the Cauchy-Schwarz or Jensen's inequalities. The core difference between the \ac{GRSST} and the MM algorithm family is that the surrogate function in \ac{GRSST} is the \ac{RMISE} function, which is minimized with respect to the chosen bandwidth in the \ac{KDE} as opposed to model parameters. A rigorous treatment of convergence is still missing in the literature but is beyond the scope of this paper. This is why we also rely on simulation studies to judge the effectiveness of the \ac{AGRSST}, just as in the original \ac{GRSST} paper \citep{gross2016}.

\subsection{The Augmented Groß–Rendtel–Schmid–Schmon–Tzavidis Estimator}
\label{AGRSST}

The precision of the \ac{GRSST} estimator is naturally constrained by the size of geographical areas which determine the sets $\left\{A_{1}, .., A_{d}, .., A_{D} \right\}$ \citep{gross2016b}. The algorithm does not require any additional information about the distribution inside the sets $A_{d}$. The \ac{AGRSST} estimator is a heuristic which enables the use of an auxiliary density $P_{\text{\tiny{AUX}}}(\boldsymbol{\ell})$ aimed at improving the \ac{GRSST} estimate $P_{\text{\tiny GRSST}}(\boldsymbol{\ell})$. Hereby, $P_{\text{\tiny{AUX}}}(\boldsymbol{\ell})$ is thought to be similar to $P_{Z}(\boldsymbol{\ell})$ due to theoretical considerations. A classical example, which we will investigate later on, is the distribution of human population $P_{Z}(\boldsymbol{\ell})$, where the distribution of nighttime light emissions is a natural candidate as an auxiliary density, because increased emissions of light may correlate with an increased human population. Both densities, $P_{\text{\tiny GRSST}}(\boldsymbol{\ell})$ and $P_{\text{\tiny AUX}}(\boldsymbol{\ell})$, can be thought of as imperfect estimations of the true distribution $P_{Z}(\boldsymbol{\ell})$. The density $P_{\text{\tiny GRSST}}(\boldsymbol{\ell})$ is imperfect because of the coarseness of the input data $X$, and $P_{\text{\tiny{AUX}}}(\boldsymbol{\ell})$ because of the generally unknown level of similarity between $P_{Z}(\boldsymbol{\ell})$ and $P_{\text{\tiny{AUX}}}(\boldsymbol{\ell})$. \Cref{alg:AGRSST} depicts the steps to obtain the \ac{AGRSST} estimator $P_{\text{\tiny{AGRSST}}}(\boldsymbol{\ell})$.

\begin{algorithm}
	\caption{Augmented Groß–Rendtel–Schmid–Schmon–Tzavidis Estimator}
	\label{alg:AGRSST}
\begin{algorithmic}[1] % The number 1 means numbered lines
	\State Run the standard GRSST algorithm to obtain $P_{\text{\tiny GRSST}}(\boldsymbol{\ell})$.
	\vspace{5pt}  % Adds vertical space between steps
	\State Evaluate $P_{\text{\tiny{AUX}}}(\boldsymbol{\ell})$ and $P_{\text{\tiny GRSST}}(\boldsymbol{\ell})$ on a fine grid $\mathbf{G}$.
	\vspace{5pt}  % Adds vertical space between steps
	\State Calculate representative mean densities of $P_{\text{\tiny{AUX}}}(\boldsymbol{\ell})$ and $P_{\text{\tiny GRSST}}(\boldsymbol{\ell})$ on the level of $A_d$ to obtain $\mathbf{m}_{\text{\tiny{AUX}}}$ and $\mathbf{m}_{\text{\tiny{GRSST}}}$.
	\vspace{5pt}  % Adds vertical space between steps
	\State Calculate the correlation $\hat \gamma = \text{cor}(\mathbf{m}_{\text{\tiny{AUX}}}, \mathbf{m}_{\text{\tiny{GRSST}}})$.
	\vspace{5pt}  % Adds vertical space between steps
	\State If $\hat \gamma < 0$, invert $P_{\text{\tiny{AUX}}}(\boldsymbol{\ell})$ on $\mathbf{G}$ by setting it to 
	$\frac{\max \left[P_{\text{\tiny{AUX}}}(\boldsymbol{\ell})\right] - P_{\text{\tiny{AUX}}}(\boldsymbol{\ell})}{\Sigma_{i} \left[P_{\text{\tiny{AUX}}}(\boldsymbol{\ell}) \right]}$.
	\vspace{5pt}  % Adds vertical space between steps
	\State Determine $P_{Z}^{\text{\tiny temp}}(\boldsymbol{\ell}) =  \hat\gamma  P_{\text{\tiny{AUX}}}(\boldsymbol{\ell}) + \left[1-   \hat \gamma  \right] P_{\text{\tiny GRSST}}(\boldsymbol{\ell})$.
	\vspace{5pt}  % Adds vertical space between steps
	\State Define the geocoordinate counts per area as $c_d = \sum_{i \in n} \mathbbm{1}\left( x_i = A_d \right)$.
	\vspace{5pt}  % Adds vertical space between steps
	\State Benchmark by sampling $c_d$ geocoordinates in each $A_d$ according to $P_Z^{\text{\tiny temp}}(\boldsymbol{\ell})$. Run a KDE with these geocoordinates to obtain a final estimate $P_{\text{\tiny{AGRSST}}}(\boldsymbol{\ell})$.
\end{algorithmic}
\end{algorithm}

The core idea of the \ac{AGRSST} estimator is to create a weighted combination of the standard \ac{GRSST} density and an auxiliary density $P_{\text{\tiny{AUX}}}(\boldsymbol{\ell})$, which on average produces a better estimate. This is because if $P_{\text{\tiny{AUX}}}(\boldsymbol{\ell})$ is unreliable, the estimate largely depends on the more robust \ac{GRSST}. However, if $P_{\text{\tiny{AUX}}}(\boldsymbol{\ell})$ is similar to the true distribution $P_Z(\boldsymbol{\ell})$, the combined estimate benefits by giving $P_{\text{\tiny{AUX}}}(\boldsymbol{\ell})$ a higher weight. In this context, $P_{\text{\tiny GRSST}}(\boldsymbol{\ell})$ serves as a baseline estimate grounded in known (population) data, while $P_{\text{\tiny{AUX}}}(\boldsymbol{\ell})$ has the potential to represent certain aspects of $P_Z(\boldsymbol{\ell})$ much better than $P_{\text{\tiny GRSST}}(\boldsymbol{\ell})$.

However, it is generally not clear how close $P_{\text{\tiny GRSST}}(\boldsymbol{\ell})$ is to $P_{Z}(\boldsymbol{\ell})$, and as a result, it is also not clear how much weight should be put on it. For this problem we propose a basic heuristic of setting the weight on $P_{\text{\tiny{AUX}}}(\boldsymbol{\ell})$ to be equivalent to $\hat \gamma = \text{cor}(\mathbf{m}_{\text{\tiny{AUX}}}, \mathbf{m}_{\text{\tiny{GRSST}}})$, which is the correlation of $P_{\text{\tiny{AUX}}}(\boldsymbol{\ell})$ and $P_{\text{\tiny GRSST}}(\boldsymbol{\ell})$ on the level of the $A_{d}$ geographical areas. The vectors $\mathbf{m}_{\text{\tiny{AUX}}}$ and $\mathbf{m}_{\text{\tiny{GRSST}}}$ are of length $D$ and contain the average densities of $P_{\text{\tiny{AUX}}}(\boldsymbol{\ell})$ and the \ac{GRSST} input data $\left(x_i\right)$. We compute $\mathbf{m}_{\text{\tiny{AUX}}}$ independently of the unit underlying the auxiliary data by taking the average of $P_{\text{\tiny{AUX}}}(\boldsymbol{\ell})$ evaluated at the gridpoints inside each $A_{d}$. Since the unit of the \ac{GRSST} input data is always geocoordinate counts we can compute $\mathbf{m}_{\text{\tiny{GRSST}}}$ by dividing the geocoordinate counts per area by the actual geographical size of their corresponding areas $A_d$, so $c_d \cdot \text{size}\left(A_d\right)$. Using linear correlation has the advantage of offering a measure that is inherently constrained within the range of $[-1, 1]$, which makes it suitable as a weight of the convex combination $P_{Z}^{\text{\tiny temp}}(\boldsymbol{\ell}) =  \hat \gamma  P_{\text{\tiny{AUX}}}(\boldsymbol{\ell}) + \left[1- \hat \gamma \right] P_{\text{\tiny GRSST}}(\boldsymbol{\ell})$. Depending on the nature of the auxiliary variable behind $P_{\text{\tiny{AUX}}}(\boldsymbol{\ell})$, the intermediate density $P_{Z}^{\text{\tiny temp}}(\boldsymbol{\ell})$ might lose some of the information contained in the \ac{GRSST} input data. To correct for such distortions, step 8 benchmarks the process so that the final estimate $P_{\text{\tiny{AGRSST}}}(\boldsymbol{\ell})$ is derived from a set of geocoordinates that align with the overall proportions specified by the \ac{GRSST} input data. In some settings, $P_{\text{\tiny{AUX}}}(\boldsymbol{\ell})$ might exhibit a negative correlation with $P_{Z}(\boldsymbol{\ell})$. For example, if $P_{\text{\tiny{AUX}}}(\boldsymbol{\ell})$ represents the density of trees, we might expect a negative relationship with $P_{Z}(\boldsymbol{\ell})$ if it represents human population density. For such cases we propose a pragmatic inversion of $P_{\text{\tiny{AUX}}}(\boldsymbol{\ell})$ as seen in step 6 of \cref{alg:AGRSST}. If $P_{\text{\tiny{AUX}}}(\boldsymbol{\ell})$ is deemed to be quite informative of $P_{Z}(\boldsymbol{\ell})$ a intuitive approach of obtaining an estimate that combines the information of the auxiliary density with the one from the \ac{GRSST} input data is to sample $c_d$ many geocoordinates in each $A_d$ weighted by $P_{\text{\tiny{AUX}}}(\boldsymbol{\ell})$. This method is, apart from the missing addition of the smoothing parameter $c$ in \Cref{alg:GRSST}'s step 4, almost equivalent to setting $\hat{\gamma} = 1$ in the \ac{AGRSST}. We decided to abstain from adding the constant $c$ in order to not disturb the information of the auxiliary density. It is a natural alternative for the \ac{AGRSST}; we refer to this estimator as \ac{AGRSST}1 and denote its density by $P_{\text{\tiny{AGRSST1}}}(\boldsymbol{\ell})$, and also investigate it.

\subsection{Obtaining Auxiliary Densities}
\label{sec:obtainaux}

In practice, $P_{\text{\tiny{AUX}}}(\boldsymbol{\ell})$ will often need to be derived from raster data measuring some physical quantity. \Cref{alg:PAUXapp} outlines our recommended method for converting this raw raster data into the $P_{\text{\tiny{AUX}}}(\boldsymbol{\ell})$ density. While it's possible to skip step (3) in \Cref{alg:PAUXapp} and directly use $w^{std}(\boldsymbol{g}_j)$ in the \ac{AGRSST}, our simulations indicate that creating $P_{\text{\tiny{AUX}}}(\boldsymbol{\ell})$ from a smoothed density, rather than the raw standardized data, generally leads to a slightly lower \ac{RMISE}. If the auxiliary data is also given by a (more granular) choropleth count map, then we recommend using the standard \ac{GRSST} to obtain $P_{\text{\tiny{AUX}}}(\boldsymbol{\ell})$.

\begin{algorithm}
	\caption{Obtaining $P_{\text{\tiny{AUX}}}(\boldsymbol{\ell})$ from raster data}
	\label{alg:PAUXapp}
	\begin{algorithmic}[1] % The number 1 means numbered lines
		\setlength{\itemsep}{5pt}
		\State Associate each grid point $\boldsymbol{g}_j$ in $\mathbf{G}$ with the raw data value $r(\boldsymbol{g}_j)$ of the corresponding raster cell it falls into.
		\State Draw a sample from $\mathbf{G}$ using $w^{std}(\boldsymbol{g}_j) = r(\boldsymbol{g}_j) \cdot \big( \sum_{\mathbf{G}} r(\boldsymbol{g}_j) \big)^{-1}$ as a sample weight. The sample size needs to be sufficiently large and can be a scaled version of the population sample size $f\cdot n$.
		\State Estimate a standard \ac{KDE} on the sampled grid points resulting in $P_{\text{\tiny{AUX}}}(\boldsymbol{\ell})$.
	\end{algorithmic}
\end{algorithm}

\section{Simulation}
\label{sec:simulation}
\subsection{Simulation Setup}
\label{sim_set}

\begin{figure}[h]
	\centering
	\includegraphics[width=1\linewidth]{"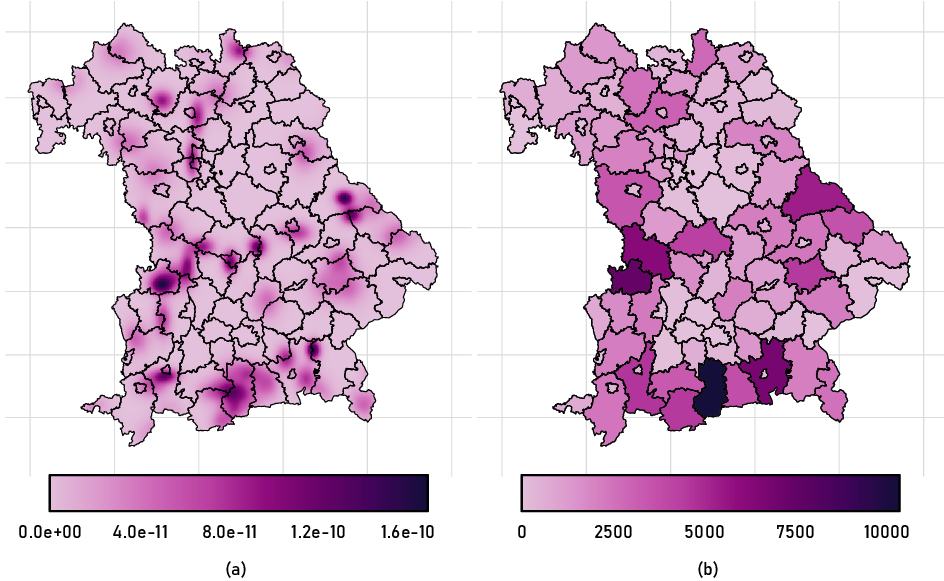"}
	\caption{(a) Exemplary artificial Bavarian population density 
		$P_{Z}^{(t)}(\boldsymbol{\ell})$  of an iteration run $t$ and (b) the corresponding \ac{GRSST} input data $\mathbf{x}^{(t)}$ resulting from  $250\,000$ draws from $P_{Z}^{(t)}(\boldsymbol{\ell})$.}
	\label{fig:1}
\end{figure}

In this section, we present the setup of the Monte Carlo simulation study used to assess the performance of the estimators introduced in \Cref{sec:methodology}. We use the map of the German state of Bavaria, which is obtained from the German \citet{federalagencyforcartographyandgeodesy2024}. The simulation is based on $t \in \{1, 2, \ldots, T = 400\}$ iterations. In each iteration, an artificial true Bavarian population density $P_{Z}(\boldsymbol{\ell})$ is simulated. $P_{Z}(\boldsymbol{\ell})$ is obtained by randomly drawing 80 times from the geocoordinates $\boldsymbol{g}_j \in \{ (\ell_{j}^{o}, \ell_{j}^{a}) \in \mathbb{R}^2 \}$ on the fine grid $\mathbf{G}$. These geocoordinates are stored in the vector $\boldsymbol{\mu}$, which is set to the location parameter of a Gaussian mixture, such that

\begin{equation}
	\label{eq:mixture}
	P_{Z}^{(t)}(\boldsymbol{\ell}) = \sum_{k=1}^{K} \pi_k \, \mathcal{N}(\mathbf{\boldsymbol{\ell}} \mid \mu_k, \boldsymbol{\Sigma}_k), \quad \sum_{k=1}^{K} \pi_k = 1, \quad 0 \leq \pi_k \leq 1, \quad \boldsymbol{\Sigma}_k = 
	\begin{bmatrix}
		\sigma_{k1}^2 & 0 \\
		0 & \sigma_{k2}^2
	\end{bmatrix}, 
\end{equation}

where the index $k \in \{1, 2, \ldots, K = 80\}$ enumerates each Gaussian component. The diagonal elements of the covariance matrices $\boldsymbol{\Sigma}_k$ are also randomly drawn, $\pi_k$ is set to $1/80$. For each simulation run, we draw an artificial Bavarian population of $n = 250\,000$ from $P_{Z}^{(t)}(\boldsymbol{\ell})$ using $\mathbf{G}$. The geographical aggregation level for the \ac{GRSST} is the German \ac{NUTS}-3 district (Kreise) level. The input vector $\mathbf{x}$ for the \ac{GRSST} is obtained by mapping the drawn geocoordinates $\mathbf{z}$ to the corresponding district $A_{d}$ in which they are located. \Cref{fig:1} shows the simulated true density $P_{Z}^{(t)}(\boldsymbol{\ell})$ and the aggregated sampled geocoordinates on district level for an exemplary simulation run $t$.

We obtain the auxiliary density $P_{\text{\tiny{AUX}}}^{(t)}(\boldsymbol{\ell})$ by taking the true density as a baseline $P_{Z}^{(t)}(\boldsymbol{\ell})$ and adding normally distributed distortions with varying variances. Generally, the \ac{GRSST} (\ac{AGRSST}) is a method designed to produce relatively reliable results for input data of low quality, due to aggregation. Depending on the target variable at hand, also $P_{\text{\tiny{AUX}}}(\boldsymbol{\ell})$ might be derived from data that is only available in a spatially aggregated form. In the simulation, we therefore base $P_{\text{\tiny{AUX}}}(\boldsymbol{\ell})$ on the German municipality (Gemeinden) level. The German municipalities are nested within the \ac{NUTS}-3 level. \Cref{alg:PAUX} depicts the process applied to obtain $P_{\text{\tiny{AUX}}}^{(t)}(\boldsymbol{\ell})$ from $P_{Z}^{(t)}(\boldsymbol{\ell})$. Note that \Cref{alg:PAUX} simply aggregates the true density, averages and standardizes it on the municipality level, draws geocoordinates according to these values, re-aggregates them on the municipality level, and applies a \ac{GRSST} estimator on these aggregates to obtain a smooth density.

\begin{algorithm}[H]
	\caption{Obtaining $P_{\text{\tiny{AUX}}}^{(t)}(\boldsymbol{\ell})$ from $P_{Z}^{(t)}(\boldsymbol{\ell})$.}
	\label{alg:PAUX}
	\begin{algorithmic}[1] % The number 1 means numbered lines
		\setlength{\itemsep}{5pt}
		\State Denote the sets of geocoordinates that fall inside the $p$-th municipality with $\left\{Q_{1}, .., Q_{p}, .., Q_{P}\right\}$.
		\State Evaluate $P_{Z}^{(t)}(\boldsymbol{\ell})$ on $\mathbf{G}$ and calculate the mean density $\mu_{Q_{p}}$ of the grid points of $\mathbf{G}$ inside every set $Q_{p}$, $\mu_{Q_{p}} = \sum_{\boldsymbol{g}_j \in Q_{p}} P_{Z}^{(t)}(\boldsymbol{g}_j) \left( \sum_{\boldsymbol{g}_j \in Q_{p}} \mathds{1}_{\boldsymbol{g}_j \in Q_{p}}(\boldsymbol{g}_j) \right)^{-1}$.
		\State Draw a sample (we chose the population size $n = 250\,000$) from $\mathbf{G}$
		using \\ $w^{std}_{\mu_{Q_{p}}}(\boldsymbol{g}_j \in Q_{p}) = \mu_{Q_{p}} \left(\sum_{\boldsymbol{g}_j \in \mathbf{G}} w_{\mu_{Q_{p}}}(\boldsymbol{g}_j \in Q_{p}) \right)^{-1}$ as a sample weight, where $w_{\mu_{Q_{p}}}(\boldsymbol{g}_j \in Q_{p}) = \mu_{Q_{p}}$.
		\vspace{5pt}
		\State Aggregate the samples drawn in step 3 to the municipality level $\{Q_{1}, \ldots, Q_{p}, \ldots, Q_{P}\}$.
		\State Run the standard \ac{GRSST} algorithm with the aggregates of the previous step, which returns $P_{\text{\tiny{AUX}}}^{(t)}(\boldsymbol{\ell})$.
	\end{algorithmic}
\end{algorithm}

 To simulate auxiliary densities of different quality, we distort the $\mu_{Q_{p}}$ with normally distributed errors with different variances.

\begin{equation}
	\label{eq:distort}
	\begin{aligned}
		\mu_{Q_{p}, \mathtt{d}} = \mu_{Q_{p}} + \epsilon_{\mathtt{d}}, \quad \epsilon_{\mathtt{d}} \sim \mathcal{N} \left(0, \left( \mathtt{d} \cdot  \mu_{Q_{p}} \cdot \left(\sum_{p = 1}^{P} \mu_{Q_{p}} \right)^{-1} \right)^{2}\right),
	\end{aligned}
\end{equation}

where the standard deviation of $\epsilon$ is set to the standardized average density over all the municipalities scaled by $\mathtt{d} \in \{0, 0.5, 1, 2.5, 5, 10, 15, 20 \}$. Replacing $\mu_{Q_{p}}$ with $\mu_{Q_{p}, \mathtt{d}}$ in step 2 of \Cref{alg:PAUX} leads to auxiliary densities of varying quality, denoted by $P_{\text{\tiny{AUX}},\mathtt{d}}^{(t)}(\boldsymbol{\ell})$.

Following \citet{gross2016}, the primary measure we use to judge the quality of an estimate $\hat P_{Z}(\boldsymbol{\ell})$ of $P_{Z}(\boldsymbol{\ell})$ is the \ac{RMISE}, which we approximate with

\begin{equation}
	\label{eq:rmise}
	\begin{aligned}
		RMISE\left(\hat P_{Z}(\boldsymbol{\ell})  \right) &= \sqrt{ \mathbb{E} \left[ \int \left( P_{Z}(\boldsymbol{\ell}) - \hat P_{Z}(\boldsymbol{\ell}) \right)^2 \, dz \right] } \approx \sqrt{ \frac{1}{m} \sum_{g \in \mathbf{G}} \left( P_{Z}(\boldsymbol{g}_j)- \hat P_{Z}(\boldsymbol{g}_j) \right)^2 \delta_{\boldsymbol{g}_j}^2 } \\ &\propto \sqrt{ \frac{1}{m} \sum_{g \in \mathbf{G}} \left( P_{Z}(\boldsymbol{g}_j)- \hat P_{Z}(\boldsymbol{g}_j) \right)^2},
	\end{aligned}
\end{equation}

the term $\delta$ refers to the side lengths of the square pixels, where the $\boldsymbol{g}_j$ are the centroids. The grid is evenly spaced, with $\delta^2_{\boldsymbol{g}_j}$ being a scalar for all $\boldsymbol{g}_j$. In the rest of this paper, we refer to \ac{RMISE} as the last term of \cref{eq:rmise}.

\subsection{Simulation Results}
\label{sim_res}

\begin{figure}[h]
	\centering
	\includegraphics[width=1\linewidth]{"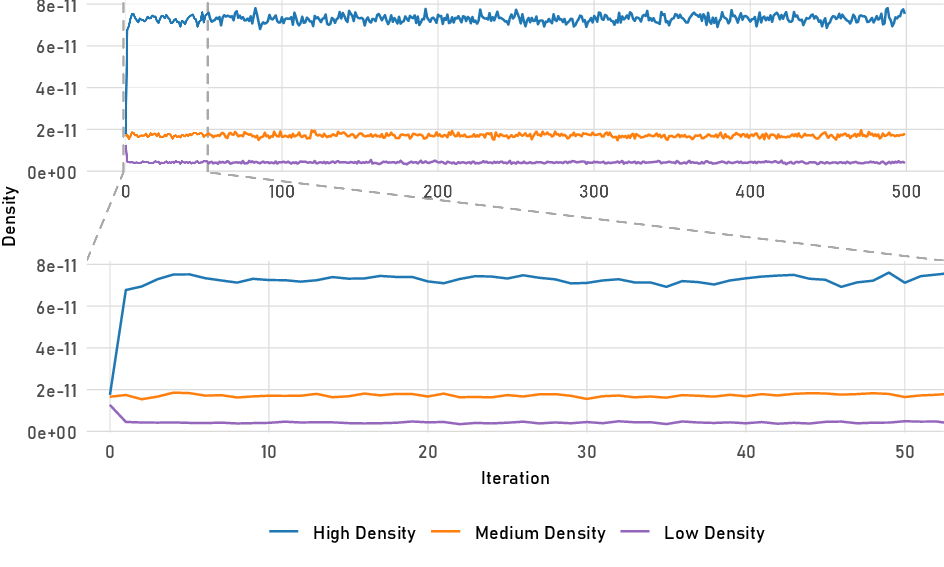"}
	\caption{The development of $\hat{P}_Z^{(l)}(\boldsymbol{\ell})$ over $M = 1000$ iterations of the \Ac{GRSST} evaluated at 3 different $\boldsymbol{g}_j$. Blue (purple, orange) represents a geocoordinate that displayed relatively high (medium, low) density in the first ($l = 1$) iteration. A zoomed-in view of the first 50 iterations is shown in the lower panel.}
	\label{fig:2}
\end{figure}

In this section, we discuss the results obtained from the Monte Carlo simulation study laid out in the previous section. For the burn-in and additional iterations, we chose $B = 30$ and $L = 20$. \Cref{fig:2} depicts the behavior of $P_{Z}^{(t)}(\boldsymbol{\ell})$ at three different locations; one that showed high (medium, low) density in the first ($l = 1$) iteration of the algorithm. It can be seen that after the pilot iteration ($l = 0$), the densities seem to quickly reach a behavior that suggests that $P_{Z}^{(t)}(\boldsymbol{\ell})$ is roaming around the stationary distribution of the Markov chain (if it exists). Additionally, choosing $B = 30$ and $L = 20$ surpasses the recommendations in the supplementary material of \citet{gross2016b}. We therefore conclude that the chosen number of $50$ iterations is sufficient in our context.

\begin{figure}[H]
	\centering
	\includegraphics[width=1\linewidth]{"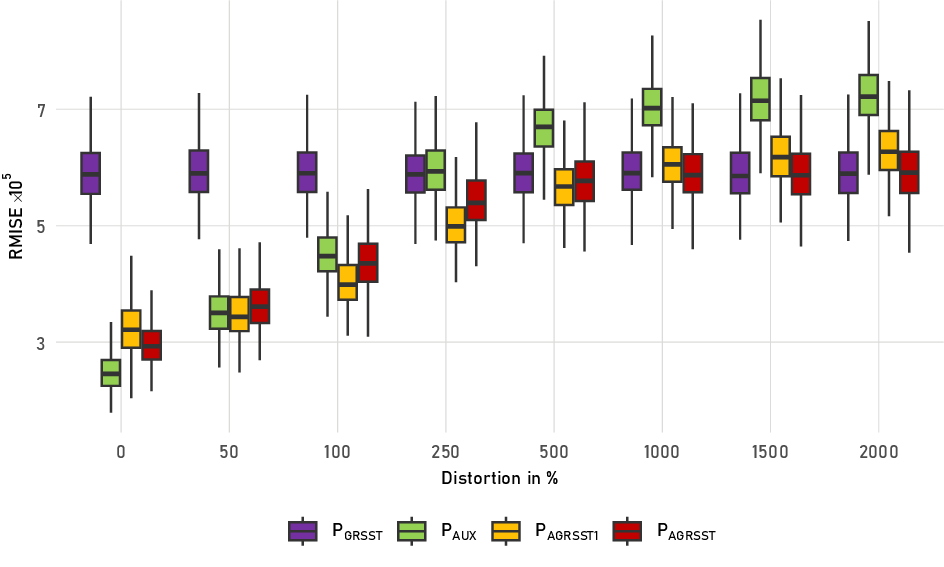"}
	\caption{Boxplots of the \ac{RMISE} of $P_{\text{\tiny{GRSST}}}(\boldsymbol{\ell})$, $P_{\text{\tiny{AUX}}}(\boldsymbol{\ell})$, $P_{\text{\tiny{AGRSST}}}(\boldsymbol{\ell})$ and $P_{\text{\text{\tiny{AGRSST1}}}}(\boldsymbol{\ell})$ over increasing levels of distortion of $P_{\text{\tiny{AUX}}}(\boldsymbol{\ell})$.}
	\label{fig:3}
\end{figure}

The boxplots in \Cref{fig:3} display the distribution of the \ac{RMISE} across 400 simulation runs for each distortion level and estimator used: $P_{\text{\tiny{GRSST}}}(\boldsymbol{\ell}), P_{\text{\tiny{AUX}}}(\boldsymbol{\ell})$, $P_{\text{\tiny{AGRSST}}}(\boldsymbol{\ell})$, and $P_{\text{\text{\tiny{AGRSST1}}}}(\boldsymbol{\ell})$. As expected, the \ac{RMISE} of the \ac{GRSST} estimator remains almost constant across different distortion levels, as it is completely independent of $P_{\text{\tiny{AUX}}}(\boldsymbol{\ell})$; minor differences can be attributed to sampling uncertainty. All estimates that utilize the auxiliary density $P_{\text{\tiny{AUX}}}(\boldsymbol{\ell})$ perform significantly better than the \ac{GRSST} for low levels of distortion. This advantage diminishes and even reverses in the case of $P_{\text{\text{\tiny{AGRSST1}}}}(\boldsymbol{\ell})$ with increasing levels of distortion. The auxiliary density $P_{\text{\tiny{AUX}}}(\boldsymbol{\ell})$ behaves as anticipated: since it is obtained via a \ac{GRSST} on the municipality level, its \ac{RMISE} is lower than that of the \ac{GRSST} on the district level. At elevated levels of distortion, $P_{\text{\tiny{AUX}}}(\boldsymbol{\ell})$ is derived from samples of a density that exhibits increasing divergence from the true density $P_Z(\boldsymbol{\ell})$. Consequently, its \ac{RMISE} is generally much higher than that of the \ac{GRSST} estimator. The poor performance of $P_{\text{\tiny{AUX}}}(\boldsymbol{\ell})$ at high distortion levels highlights how, in practice, simply choosing a seemingly similar density as an estimate for another should be avoided. 

The methods that utilize the \ac{GRSST} input data $\big(P_{\text{\tiny{AGRSST}}}(\boldsymbol{\ell}) \text{ and } P_{\text{\text{\tiny{AGRSST1}}}}(\boldsymbol{\ell})\big)$ mitigate the effects of a very poor auxiliary density. The comparison between $P_{\text{\tiny{AGRSST}}}(\boldsymbol{\ell})$ and $P_{\text{\text{\tiny{AGRSST1}}}}(\boldsymbol{\ell})$ reveals lower \ac{RMISE} values for the \ac{AGRSST} at both the lowest and high distortion levels, whereas $P_{\text{\text{\tiny{AGRSST1}}}}(\boldsymbol{\ell})$ outperforms the \ac{AGRSST} for moderate distortions. For highly informative auxiliary densities where the only source of divergence from the true density is the averaging at the municipality level (distortion = 0), the benchmarking step in $P_{\text{\tiny{AGRSST}}}(\boldsymbol{\ell})$ and $P_{\text{\text{\tiny{AGRSST1}}}}(\boldsymbol{\ell})$ does not improve the estimate but increases it due to the noise from the additional \ac{KDE} process. In practice, this effect would only be relevant for auxiliary densities that are very similar to the true density and is therefore of little practical consequence. The \ac{AGRSST}'s advantage over the \ac{AGRSST}1 at zero distortion likely stems from its ability to assign a small weight to the \ac{GRSST} estimate, which slightly smooths the municipality-level based $P_{\text{\tiny{AUX}}}(\boldsymbol{\ell})$. In the case of moderate distortions (100 - 500), the additional information about the true density incorporated into $P_{\text{\text{\tiny{AGRSST1}}}}(\boldsymbol{\ell})$ outweighs this advantage, and $P_{\text{\text{\tiny{AGRSST1}}}}(\boldsymbol{\ell})$ outperforms $P_{\text{\tiny{AGRSST}}}(\boldsymbol{\ell})$. For very poor auxiliary densities, \ac{AGRSST} performs better than $P_{\text{\text{\tiny{AGRSST1}}}}(\boldsymbol{\ell})$ due to its ability to assign weight to the \ac{GRSST} estimate, which is superior and more conservative because of its independence from the (heavily distorted) auxiliary density.

\begin{figure}[H]
	\centering
	\includegraphics[width=1\linewidth]{"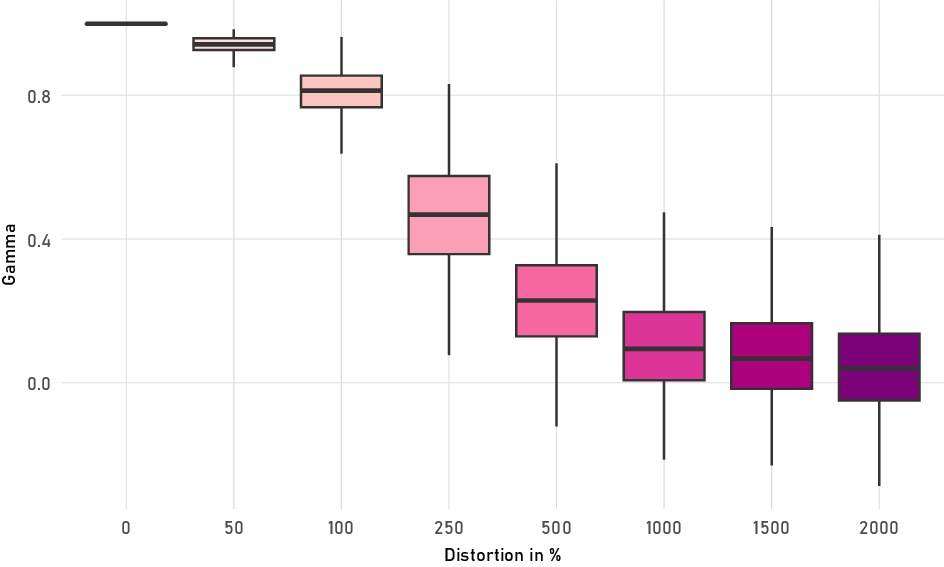"}
	\caption{Boxplots of the calculated value of $\hat \gamma$ over increasing levels of distortion of $P_{\text{\tiny{AUX}}}(\boldsymbol{\ell})$.}
	\label{fig:4}
\end{figure}

The above described behavior is confirmed by \Cref{fig:4}, which shows the dependence of $\hat \gamma$ on the distortions. We deem the displayed relationship between $\hat \gamma$ and the distortion levels as generally desirable, because it drives the advantages of the \ac{AGRSST} illustrated in \Cref{fig:3}. Though, when examining the interplay between the true density $P_Z(\boldsymbol{\ell})$, $P_{\text{\tiny{AUX}}}(\boldsymbol{\ell})$ and $\hat \gamma$ the limitations of the \ac{AGRSST} become apparent. The performance of the \ac{AGRSST} estimator relies on the fact that $\hat \gamma$ properly reflects the similarity between the true- and the auxiliary density. Confounding factors that can negatively impact the reflection of this relationship via $\hat \gamma$ can be:
\begin{itemize}
	\item Nonlinearity: $P_Z(\boldsymbol{\ell})$ and $P_{\text{\tiny{AUX}}}(\boldsymbol{\ell})$ might be related in a nonlinear fashion. Since $\hat \gamma$ is essentially a standard empirical correlation, it might not, or only partly, capture such relationships.
	\item Ecological fallacy \citep{holt1996}: Since all the information about $P_Z(\boldsymbol{\ell})$ is limited to the \ac{GRSST} input data at its respective geographical aggregation level, we calculate $\hat \gamma$ at that level. The \ac{AGRSST} assumes, though, that $\hat \gamma$ is a valid similarity measure at the level of the densities.
\end{itemize}

The main tool to counteract problems caused by a $\hat \gamma$ that does not reflect the quality of $P_{\text{\tiny{AUX}}}(\boldsymbol{\ell})$ is the benchmarking (step 8 in \Cref{alg:AGRSST}). It limits the amount of possible divergence of $P_{\text{\tiny{AGRSST}}}(\boldsymbol{\ell})$ from $P_{\text{\tiny{GRSST}}}(\boldsymbol{\ell})$. The worst-case scenario for the \ac{AGRSST} is a highly divergent auxiliary density from the true density, which is still highly correlated with the true density at the \ac{GRSST} input level. This scenario is covered by our simulation study; at the highest distortion levels, $P_{\text{\text{\tiny{AGRSST1}}}}(\boldsymbol{\ell})$ puts maximum weight on very poor auxiliary densities ($\hat \gamma = 1$). We observe that $P_{\text{\text{\tiny{AGRSST1}}}}(\boldsymbol{\ell})$ performs worse than the \ac{GRSST}, but the benchmarking step still ensures that divergence is limited. The \ac{AGRSST} is basically equivalent to the \ac{GRSST} at these high distortion levels. The main insights from our simulation study are that:

\begin{itemize}
	\item $P_{\text{\text{\tiny{AGRSST1}}}}(\boldsymbol{\ell})$, which benchmarks a plausible auxiliary density with the \ac{GRSST} input $c_d$, is an attractive alternative to the \ac{GRSST} with large potential gains in \ac{RMISE} but a potentially worse \ac{RMISE} for poor auxiliary densities.
	\item In contrast to $P_{\text{\text{\tiny{AGRSST1}}}}(\boldsymbol{\ell})$, the \ac{AGRSST} provides another layer of security to practitioners by limiting the weight placed on the auxiliary density when low correlation at the \ac{GRSST} input level is detected. The trade-off for this increased security is a lower \ac{RMISE} performance in the case of very informative auxiliary densities.
\end{itemize}

\section{Evaluation Study on Population Data}
\label{sec:evaluation}	
	
\subsection{Target Variable}

\begin{figure}[h]
	\centering
	\includegraphics[width=1\linewidth]{"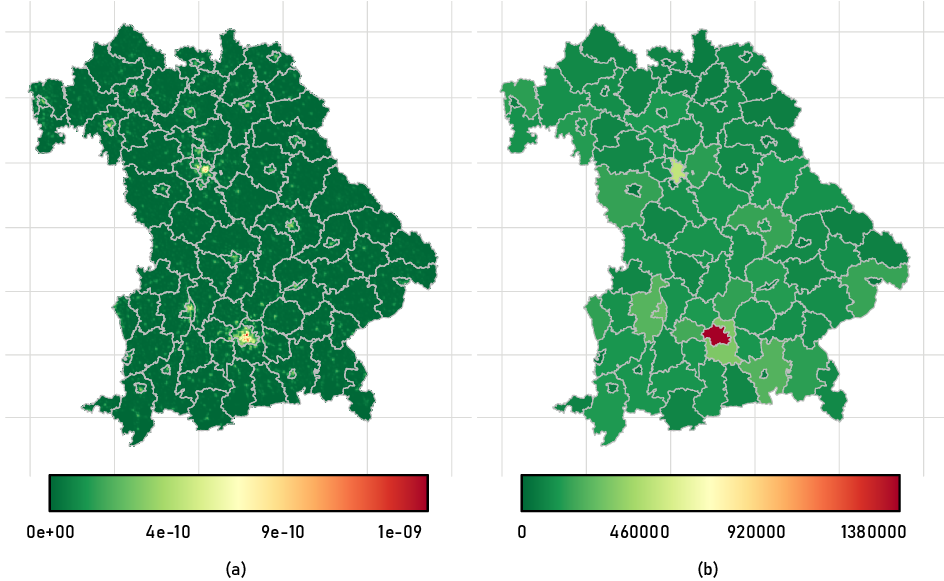"}
	\caption{(a) The assumed ground truth, a \ac{KDE} of the true Bavarian population based on the 2022 German Census 100 meter raster data and (b) the Bavarian population on the \ac{NUTS}-3 district (Kreise) level, which serve as the \ac{GRSST} input data.}
	\label{fig:5}
\end{figure}

The estimation of human population distributions is a classic area of interest in research in order to allow for targeted social- and environmental planning and policy development. Multiple contributions to the topic are, for example, added by the WorldPop research group; see, e.g., \citet{stevens2015} or \citet{sorichetta2015}. In our first application, we use recently published data from the German Census 2022 that is conducted by the \ac{Destatis}. Population counts are available in 100-meter raster cells \citep{destatis2024a} and also at the district level \citep{destatis2024}. We consider the raster data in relation to the size of Bavaria as granular enough to estimate a density $P_Z(\boldsymbol{\ell})$, which we assume to be the true population distribution. This density is estimated using a \ac{KDE} based on geocoordinates located at the centroids of each raster. So, for a raster XYZ in which X-many people live, we place X-many geocoordinates at the centroid of XYZ. This is done for all rasters in Bavaria. The true distribution $P_Z(\boldsymbol{\ell})$ follows from a \ac{KDE} on the resulting dataset (\Cref{fig:5} panel (a)). We obtain $P_{\text{\tiny{GRSST}}}(\boldsymbol{\ell})$ by running the \ac{GRSST} estimator on the district-level population data (\Cref{fig:5} panel (b)). Since highly accurate data on the population distribution is already available, the principal objective of our first application is not to obtain an even more accurate estimate of the population density, but to use and compare our methods in a real-world scenario in which we also have access to a fairly accurate estimate of the ground truth $P_Z(\boldsymbol{\ell})$.

\subsection{Auxiliary Variable}

In practice, the auxiliary density $P_{\text{\tiny{AUX}}}(\boldsymbol{\ell})$ can be derived from various sources with differing underlying units and aggregation levels. For example, in the case of nighttime light satellite raster data \citep{elvidge2021} that we use, the resolution of 15 arc-seconds ($\approx$ 300 meters at the latitude of Bavaria's geographical center) is fairly granular with respect to the sizes of the \ac{NUTS}-3 districts. In this example, the radiance of the light is measured in watt per steradian per square meter $\left(W \cdot sr^{-1} \cdot m^{-2}\right)$. For our application, we use the 2022 masked median radiance from monthly averages. The process to obtain these values includes multiple steps to adjust for distortions such as biomass burning or aurora and is documented in \citet{elvidge2021}. The general relationship between nighttime lights and population density is well established and, e.g., studied by \citet{sutton1997}, \citet{elvidge1999}, \citet{sutton2001}, \citet{wu2023}.

\begin{figure}[h]
	\centering
	\includegraphics[width=1\linewidth]{"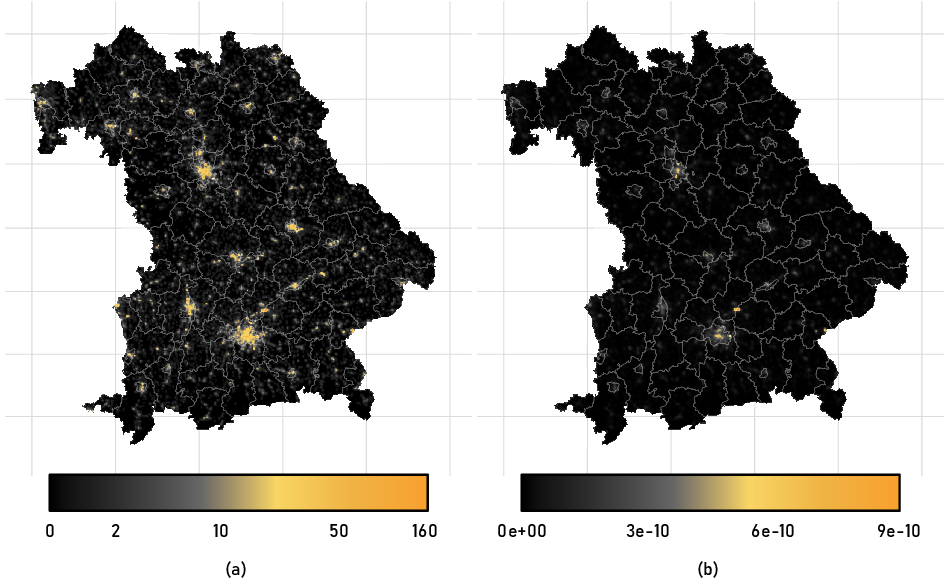"}
	\caption{(a) The raw nighttime light satellite raster data, displayed on a pseudo log-transformed color scale for better contrast, and (b) the auxiliary density $P_{\text{\tiny{AUX}}}(\boldsymbol{\ell})$ derived by \Cref{alg:PAUXapp}. The auxiliary density is also pseudo log-transformed, and values above $9 \cdot 10^{-9}$ (located at Munich airport) are cut for improved visibility.}
	\label{fig:6}
\end{figure}

In the scope of Bavaria, we deem the resolution of nighttime light radiance data granular enough to directly convert it into a density using \Cref{alg:PAUXapp}. Depending on the target variable at hand, the analyst might not have access to auxiliary data as granular as the nighttime light satellite data. In such cases, we recommend using a \ac{GRSST} for the conversion, similar to \Cref{alg:PAUX}. As the sample size in \Cref{alg:PAUXapp}, we chose the unscaled Bavarian population $(n = 13\,038\,724)$. Note that the analyst can scale this number, if computational resources allow, to further reduce the sampling uncertainty, although the gains are likely to be relatively small given the already large sample size. The smoothing effect by the \ac{KDE} can be judged visually by comparing the (transformed) raw data to $P_{\text{\tiny{AUX}}}(\boldsymbol{\ell})$ as in \Cref{fig:6}. If core features of the raw data are smoothed out, we recommend using the raw data or increasing $n$. The benchmarked auxiliary density $P_{\text{\tiny{AGRSST1}}}(\boldsymbol{\ell})$ is analogously obtained from $P_{\text{\tiny{AUX}}}(\boldsymbol{\ell})$ as in the simulation study.

\subsection{Evaluation Results}
The value for $\gamma$ is calculated as laid out in \Cref{AGRSST} and is $\hat{\gamma} \approx 0.945$, which means that the light density values are highly correlated with the Bavarian population densities at the district level. The availability of the true density allows us to determine \ac{RMISE} values, which are depicted in \Cref{tab:rmise_values}.

\color{black}

\begin{table}[htbp]
	\centering
	\begin{tabularx}{0.5\textwidth}{l>{\raggedleft\arraybackslash}X}
		\hline
		\textbf{Method} & \textbf{RMISE $\cdot 10^{5}$} \\
		\hline
		\hline
		$P_{\text{\tiny{GRSST}}}(\boldsymbol{\ell})$ & 8.3252 \\
		$P_{\text{\text{\tiny{AGRSST1}}}}(\boldsymbol{\ell})$ & 8.1803 \\
		$P_{\text{\tiny{AGRSST}}}(\boldsymbol{\ell})$ & \textbf{8.1396} \\
		$P_{\text{\tiny{AUX}}}(\boldsymbol{\ell})$ & 8.8786 \\
		\hline
		\hline
	\end{tabularx}
	\caption{\ac{RMISE} values of the different population density estimates, calculated based on census population data.}
	\label{tab:rmise_values}
\end{table}

\begin{figure}[h]
	\centering
	\includegraphics[width=1\linewidth]{"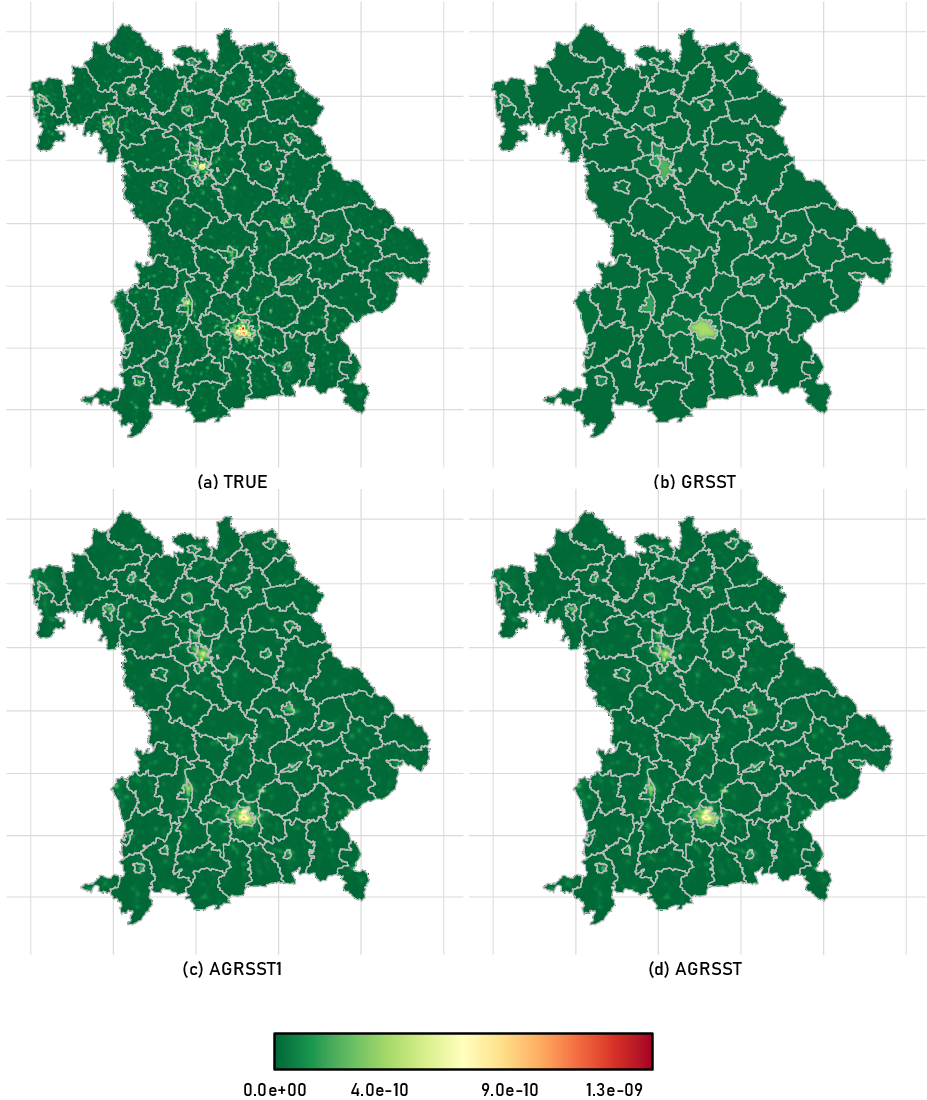"}
	\caption{(a) The assumed ground truth, a \ac{KDE} of the true Bavarian population based on the 2022 German Census 100-meter raster data, (b) the \ac{GRSST}- , (c) \ac{AGRSST}- and (d) the \ac{AGRSST}1 estimate.}
	\label{fig:7}
\end{figure}

\begin{figure}[h]
	\centering
	\includegraphics[width=1\linewidth]{"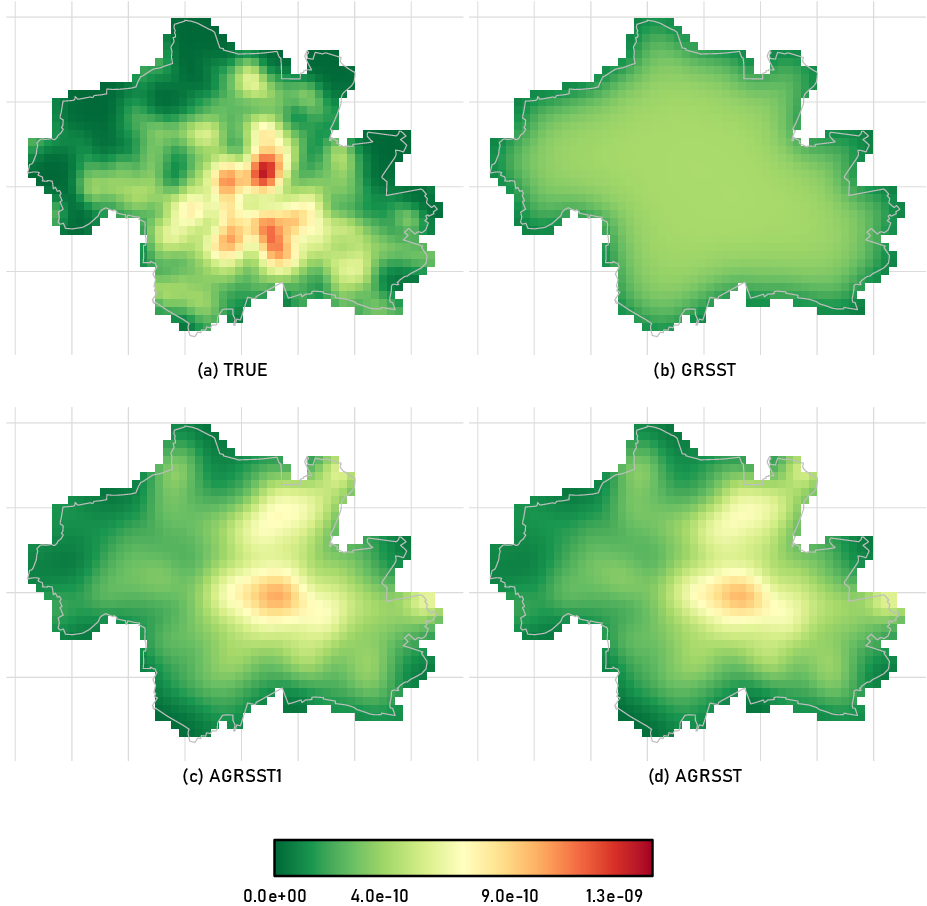"}
	\caption{(a) The assumed ground truth, a \ac{KDE} of Munich's true population based on the 2022 German Census 100-meter raster data, (b) the \ac{GRSST}- , (c) \ac{AGRSST}- and (d) the \ac{AGRSST}1 estimate.}
	\label{fig:7.1}
\end{figure}

Compared to the simulation study, the improvements of the \ac{AGRSST} over the \ac{GRSST} in this real-world example appear rather small, with a \ac{RMISE} reduction of $2.23\%$. A potential reason for such small gains could be large deviations between light intensity and actual population in areas such as airports or the at Theresienwiese in Munich (\Cref{fig:7.1}). Interestingly, the \ac{AGRSST}1 performs slightly worse than the \ac{AGRSST}; we attribute this to the additional smoothing that the mixture provides, especially in said areas with large deviations between light and population density, a dynamic that we also observed in our simulation study at the zero distortion level. The non-benchmarked $P_{\text{\tiny{AUX}}}(\boldsymbol{\ell})$, in turn, has a $6.65\%$ higher \ac{RMISE} than the \ac{GRSST}. This structure can be compared to the distortion levels of $250$ and $500$ in our simulation study, where the \ac{AGRSST} and \ac{AGRSST}1 outperform the \ac{GRSST}, even with an auxiliary density that itself has a higher \ac{RMISE} than the \ac{GRSST}. \Cref{fig:7} depicts the resulting densities; upon visual inspection, the large-scale population density seems to be much better captured by \ac{AGRSST} and \ac{AGRSST}1, contrary to the small improvement in \ac{RMISE}.

\Cref{fig:7.1}, an enlarged view of \Cref{fig:7} focusing on Munich, illustrates the dynamics of all three methods. While the \ac{GRSST} estimate appears largely uniform within Munich's boundaries, \ac{AGRSST}1 and \ac{AGRSST} manage to capture some of the true population dynamics, although both incorrectly depict the Theresienwiese area in central Munich – the site of the famous Oktoberfest – as having the highest density, rather than a low population. Given the relatively spiky nature of the auxiliary density, the more conservative nature of \ac{AGRSST} compared to \ac{AGRSST}1 is evident in its less spiky density. This example highlights the significant dependence of the \ac{AGRSST} on the auxiliary density and underscores the importance of selecting these densities based on expert subject knowledge. Despite the generally limited data availability typically associated with \ac{GRSST} usage, our overall evaluation study demonstrates the potential of \ac{AGRSST} to enhance \ac{GRSST} density estimates, both in terms of \ac{RMISE} and visual representation.

\section{Application Rabbit Population Control}

\label{sec:application}	

\subsection{Target Variable}

The Lepus europaeus, often called the brown hare or European hare, is a species of Lepus that is native to many European countries but has also been introduced into areas all over the world. \citet{bock2020} provides a comprehensive overview and literature collection on the general characteristics, distribution, ontogeny, reproduction, and other aspects of the mammal. Since the 1960s, there has been a steep decline in brown hare populations across Europe. The primary causes of this decline are believed to be the loss of landscape diversity and the emergence of diseases (see e.g., \citet{smith2005}). Since 2015, the population of brown hares in Lower Saxony has shown signs of recovery, which is likely attributable to the dry conditions experienced in recent years, which have resulted in a decline in infection rates. The implementation of local initiatives by hunters, including the creation of wildflower strips and increased predator hunting, is another potential contributing factor to the recovery \citep{graber2024}. \citet{sliwinski2019} studied the influence of different habitat variables on brown hare population density in Lower Saxony, using \acp{GAMM} fit at the municipality level. We focus on identifying areas of high and low hunting intensity. To achieve this, we estimate a density based on the number of brown hares killed per district in Lower Saxony and Bremen during the hunting season 2023/2024 (01.04.2023 - 31.03.2024) \citep{dammann-tamke2025}. The number of animals killed by hunting and/or traffic-related accidents is also referred to as the hunting bag or hunting bag statistic. While hunting bag statistics have been used as a proxy to infer trends in wildlife population densities, this practice is not always advisable for reasons such as temporary hunting bans \citep{graber2024}. As a complement to wildlife population densities, hunting bag statistics are still an important analytical tool used to determine overall variations in wildlife numbers (see e.g., \citet{havet2002}) and are needed to implement harvest management schemes \citep{aubry2020}.

\subsection{Auxiliary Variable}

\begin{figure}[h]
	\centering
	\includegraphics[width=1\linewidth]{"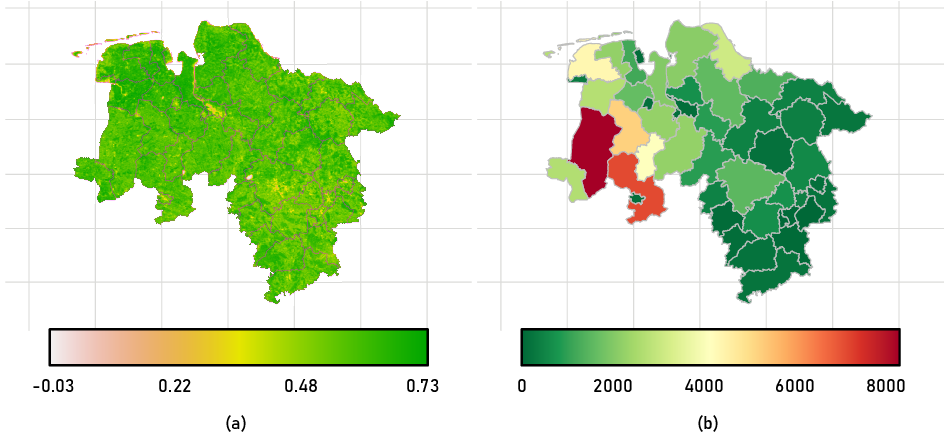"}
	\caption{(a) The \ac{NDVI} and (b) brown hare hunting bag data of Lower Saxony during the hunting season 23/24. The borders display the German \ac{NUTS}-3 district (Kreise) level.}
	\label{fig:8}
\end{figure}

We utilize the \ac{NDVI} as the underlying data for the purpose of obtaining an auxiliary density for the hunting bag data. The fundamental concept (see \citet{huete2002}) underpinning vegetation indices, such as the \ac{NDVI}, is based on the study of the spectral reflectance of leaves. The reflected radiation in the red band from leaves is relatively low due to the absorption of light by photosynthetically active pigments, such as chlorophyll. In contrast, a significant portion of \ac{NIR} is reflected, primarily due to the internal structure of plant leaves. This contrast between red and near-infrared reflectance allows for the construction of the \ac{NDVI} as a sensitive index used to quantify vegetation, given by
\begin{equation}
	NDVI = \frac{\rho_{\text{\tiny NIR}} - \rho_{\text{\tiny RED}}}{\rho_{\text{\tiny NIR}} + \rho_{\text{\tiny RED}}}.
\end{equation}
where $\rho_{\text{NIR}}$ and $\rho_{\text{RED}}$ are the surface bidirectional reflectance factors in the near-infrared and red bands. The raw data is collected by the \ac{MODIS} on board the Earth Observing System-Terra platform \citep{didan2019}. We use the \texttt{MOD13A1: 16-day 500m VI} data product \citep{didan2021}, which provides 500-meter \ac{NDVI} raster data in 16-day intervals. The specified interval (01.04.2023 - 31.03.2024) encompasses a total of 23 timestamps for raster layers. The mean of these raster values is calculated, resulting in a single \ac{NDVI} raster layer for the designated hunting season (2023/2024). The \ac{NDVI} is standardized between -1 and 1, with the majority of observed values falling between 0 (water or bare soil) and 0.8 (dense vegetation). Pane (a) in \Cref{fig:8} displays the \ac{NDVI} raw data over Lower Saxony. There is a rich body of literature (e.g., \citet{sliwinski2019}, \citet{schai-braun2012}, \citet{smith2005}) on brown hare habitat preferences, which indicates that these depend on factors such as season, temperature, precipitation, and brown hare density itself. Generally, we can conclude that brown hares prefer heterogeneous agricultural landscapes with a mix of crops, grasslands, and semi-natural habitats, avoiding large monocultures, urban areas, and intensively farmed regions. The \ac{NDVI} does not directly reflect all of these factors, such as agricultural heterogeneity, but the general assumption of a positive relationship between vegetation, expressed by \ac{NDVI}, and the brown hare hunting bag seems plausible.

\subsection{Application Results}

\begin{figure}[h]
	\centering
	\includegraphics[width=1\linewidth]{"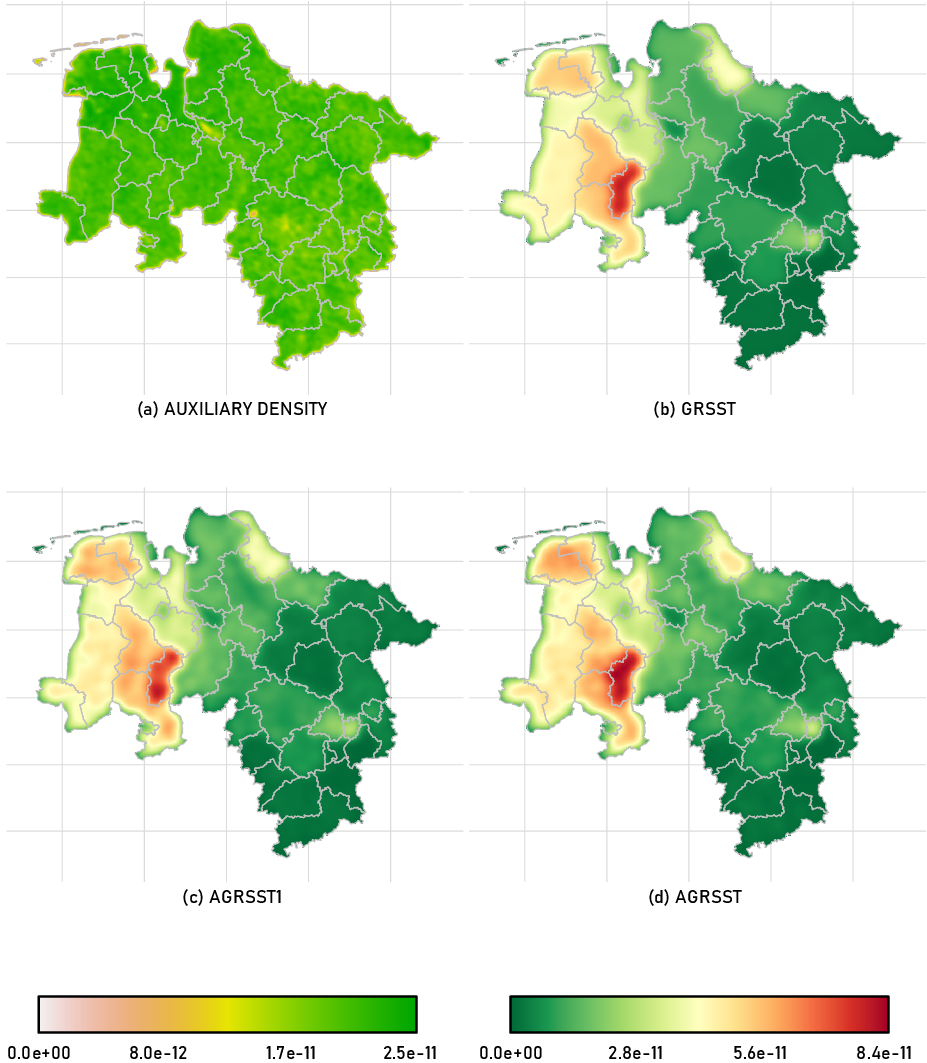"}
	\caption{(a) The auxiliary density, obtained by applying \Cref{alg:PAUXapp} on the raw \ac{NDVI} data (the corresponding color scale is on the bottom left), (b) the \ac{GRSST}, (c) \ac{AGRSST}, and (d) the \ac{AGRSST}1 estimate (their color scale is on the bottom right).}
	\label{fig:11}
\end{figure}

\begin{figure}[h]
	\centering
	\includegraphics[width=1\linewidth]{"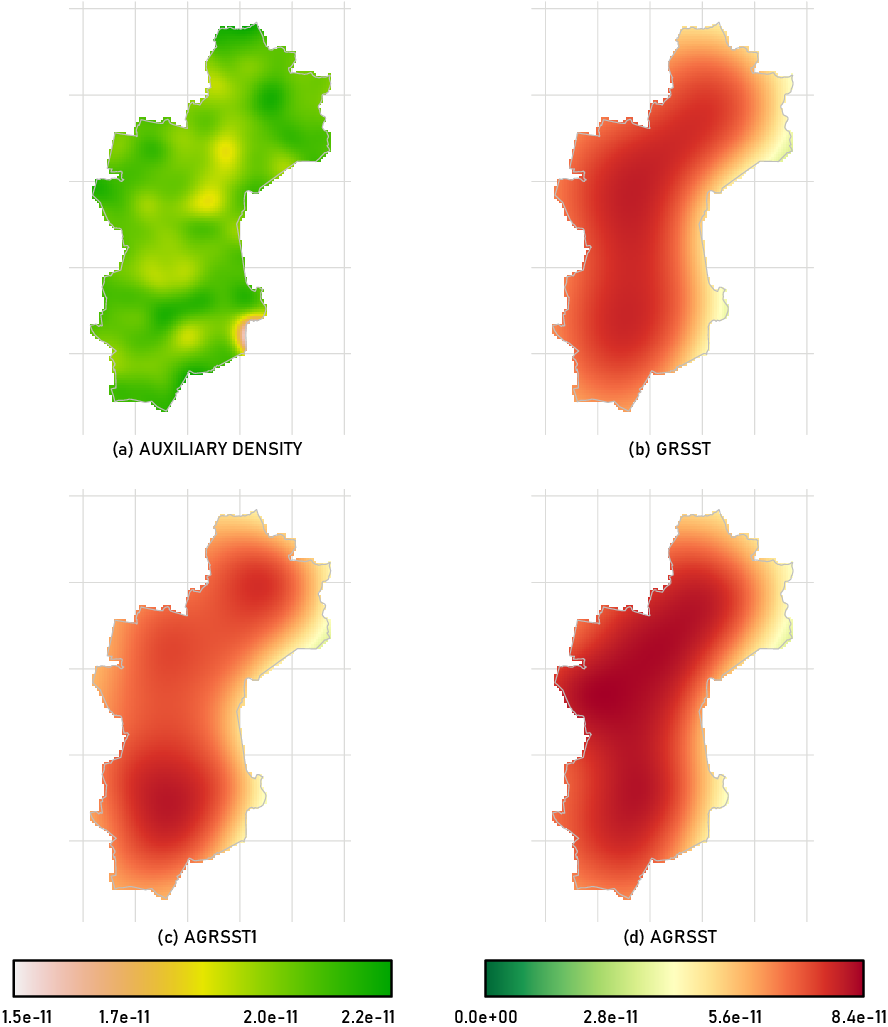"}
	\caption{A version of \Cref{fig:11}, zoomed-in on the district of Vechta. (a) The auxiliary density, obtained by applying \Cref{alg:PAUXapp} on the raw \ac{NDVI} data (the corresponding color scale is on the bottom left), (b) the \ac{GRSST}, (c) \ac{AGRSST}, and (d) the \ac{AGRSST}1 estimate (their color scale is on the bottom right).}
	\label{fig:12}
\end{figure}

Applying the steps outlined in \Cref{AGRSST} again yields a value of $\hat{\gamma} \approx 0.2067$, indicating a rather low, yet still significant correlation between \ac{NDVI} and the brown hare hunting bag. The resulting density estimates are displayed in \Cref{fig:11}. Unlike our evaluation study in the preceding section, it is not possible to determine \ac{RMISE} values here because the brown hare hunting bag data is only available at the district level. Panel (a) of \Cref{fig:11} illustrates the auxiliary density $P_{\text{\tiny{AUX}}}(\boldsymbol{\ell})$ derived from the raw \ac{NDVI} data. It is important to note that panel (a) has its own distinct color scale and legend, whereas panels (b)-(d) share a common color scale to facilitate comparisons. The figure reveals that the auxiliary density is relatively flat compared to the \ac{GRSST}. From the global perspective provided by \Cref{fig:11}, we can observe the typical behavior of \ac{GRSST}, exhibiting a rather smooth density within districts. In contrast, \ac{AGRSST}1 and \ac{AGRSST} display more structure within these districts. All three estimators indicate a west-to-east gradient with high bag densities in the west and declining densities to the east. In contrast to the absolute bag data (panel (b) in \Cref{fig:8}), the \ac{KDE}-based estimators identify a cluster of brown hare hunting in the three districts of Vechta, Cloppenburg, and Osnabrück, with Vechta being the district exhibiting by far the highest bag density. \Cref{fig:12} zooms in on the district of Vechta to better visualize the discrepancies between the three estimates; the color scales remain consistent.

The \ac{AGRSST}1 significantly reflects some of the structure of the auxiliary density shown in panel (a); the dark green high vegetation hotspots north and south of Vechta can be clearly identified. Note that we observed a much larger impact of the auxiliary density in our evaluation study. The limited effect of the \ac{NDVI} is due to two characteristics of the input data. First, the total count of $n = 64\,779$ causes the determinants of the bandwidth matrices $\mathbf{H}$ to be relatively large compared to the setting in our evaluation study, which in turn causes the resulting density estimates to be smoother. Second, the \ac{NDVI} auxiliary density is generally much smoother than the \ac{GRSST} density, which is quickly confirmed by comparing the maximum values: $2.5 \cdot 10^{-11}$ for the auxiliary density versus $8.0 \cdot 10^{-11}$ for the \ac{GRSST}. Also, contrary to our expectation, the \ac{AGRSST} estimate seems to have higher maximum density values than the \ac{AGRSST}1 and the \ac{GRSST}. Two factors contribute to this: first, the \ac{AGRSST}'s derivation without the additional constant in step 3 of \Cref{alg:GRSST} results in a spikier estimate compared to the \ac{GRSST}; second, the relatively flat auxiliary density leads to a flatter \ac{AGRSST}1 estimate. Considering these aspects, we prefer the \ac{AGRSST}1 approach for this particular application, as it is sensitive to the \ac{NDVI} input data while maintaining a high degree of smoothness. This balance already helps limit potential errors compared to the \ac{GRSST}, even without relying on the \ac{AGRSST} rationale.

\section{Conclusion}
\label{sec:conclusion}

In this paper, we propose an extension to the \ac{GRSST} estimator, which we name \ac{AGRSST}. The \ac{GRSST} is a method that broadens the applicability of \ac{KDE} to challenging scenarios involving aggregated data, enabling more profound insights compared to basic choropleth maps displaying absolute counts. These advantages include facilitating the identification of spatial clusters and enabling the transformation of area aggregates between non-hierarchical geographical systems. However, the inherent accuracy of the \ac{GRSST} remains limited by the relative geographical size of its input areas, as no direct information about the density distribution within these areas is provided. \\
The \ac{AGRSST} estimator linearly combines the auxiliary and \ac{GRSST} densities. The weight assigned to the auxiliary density is determined by its correlation with the \ac{GRSST}-derived density at the level of the aggregated spatial units. This allows the \ac{AGRSST} to adapt to the informativeness of the auxiliary data, placing more emphasis on it when it exhibits a strong positive correlation and relying more on the conservative \ac{GRSST} estimate when the correlation is weak. Furthermore, a benchmarking step ensures that the final density estimate is based on the original aggregated counts. We also highlighted a special version of the \ac{AGRSST}, the \ac{AGRSST}1, which directly samples within the aggregated units according to the auxiliary density. \\
To evaluate the effectiveness of the \ac{AGRSST} estimator, we conducted simulation study using the administrative map of Bavaria, Germany. We simulated various true population densities and generated auxiliary densities of varying quality by introducing controlled levels of distortion. The \ac{RMISE} was used as the primary metric to assess the performance of the standard \ac{GRSST} and the proposed \ac{AGRSST} variants across these different distortion levels. The results of our simulation study show that depending on the quality of the auxiliary density large accuracy improvements are possible when using the \ac{AGRSST}. It also became apparent that the correlation based convex combination serves as a mechanism which limits the negative influences of poor auxiliary densities. When fully weighting the auxiliary density (\ac{AGRSST}1), we observe the potential for greater improvements over the \ac{GRSST}, but also the risk of worse estimates in terms of \ac{RMISE}, contingent on the quality of the auxiliary density. These findings are corroborated in a real-world evaluation using German Census data and remotely sensed nighttime lights, where, owing to the high correlation between nighttime lights and population density, \ac{AGRSST}1 outperforms \ac{AGRSST}, which in turn outperforms the \ac{GRSST}.

Lastly, applying \ac{AGRSST} and \ac{AGRSST}1 to Lower Saxony's brown hare hunting bag data reveals a west-to-east density gradient, with intense hunting clustered in the Vechta, Cloppenburg, and Osnabrück districts. Comparing the estimators, \ac{AGRSST} is spikier than \ac{GRSST}, while only \ac{AGRSST}1 effectively integrates information from the \ac{NDVI} data. Considering these factors, we conclude that \ac{AGRSST}1 is the preferred method for this application.

The findings of this paper open several interesting avenues for future research in this domain. A potentially powerful enhancement could involve the integration of auxiliary densities within each iteration of the \ac{GRSST} algorithm. Indeed, during the course of this research, we explored one such approach, wherein the standard \ac{KDE} step in the \ac{GRSST} was replaced by weighted \ac{KDE} (see, e.g., \citet{wang2007}), with weights determined by the corresponding values of the auxiliary density at the locations of the input geocoordinates. Unfortunately, this specific implementation did not yield significant improvements over the \ac{AGRSST} presented herein and was thus omitted for the sake of brevity. Another promising direction for future research concerns the incorporation of multiple auxiliary densities. While the optimal strategy for weighting these distinct densities to derive a final estimate remains an open question, a potential initial approach could involve formulating an overall auxiliary density as a linear convex combination of several individual auxiliary densities, with weights potentially determined through a multi-dimensional grid search aimed at maximizing $\hat \gamma$, the estimated correlation. However, we also note that a significantly improved (auxiliary) data situation enables more sophisticated methodologies such as random forest-based dasymetric mapping \citep{stevens2015}. This generally raises questions regarding the relationship between \ac{AGRSST} and such methods, including when one approach should be preferred over the other, and what potential combinations might be beneficial. Furthermore, future work with significant practical implications includes the extension of the \texttt{Kernelheaping} package to incorporate the \ac{AGRSST} methodology, as well as the introduction of refinement schemes already established for the \ac{GRSST}. These include the boundary correction technique and the capacity to exclude (uninhabited) areas known to have a density of zero from the estimation process (\citet{erfurth2022}).

Overall, our research demonstrates the considerable potential that auxiliary densities offer in the context of density estimation with aggregated data. We recommend that practitioners consider using \ac{AGRSST} when the emphasis is on limiting potential adverse effects from incorporating an auxiliary density, and \ac{AGRSST}1 when the focus is on maximizing potential accuracy. In any case, considering the generally poor initial data situation, which largely prohibits advanced statistical modeling and uncertainty quantification techniques such as cross-validation, a sound theoretical justification for the candidate auxiliary density is indispensable. \\

\section*{Acknowledgements}

The generative AI tools ChatGPT-4o, Gemini 2.0 Flash, and Claude 3.7 Sonnet were used to improve the language, grammar, and structure of this submission. They were also employed as search engines for research purposes. The core ideas, as well as the creative and critical thinking behind this work, are solely attributed to the authors.
	
\section*{Declarations}
	
\subsection*{Funding and/or Conflicts of interests/Competing interests}
	
All authors declare no conflicts of interest.
	
\subsection*{Code Availability Statement}
	
The code used in this study is available upon request from the corresponding author.
	
\section*{Data Statement}
The hunting bag statistics for Bremen and Lower Saxony are available in the Handbook of the German Hunting Association (DJV), whose president is \citet{dammann-tamke2025}, and can be purchased online. All other datasets used are freely available under their corresponding citations.

\newpage

\setlength{\emergencystretch}{3em}

\printbibliography
	
\end{document}